\documentclass[a4paper,10pt,twocolumn]{article}
\pdfoutput=1
\usepackage{fullpage}
\usepackage{todonotes} 
\usepackage{hyperref}  
\usepackage[font=small,labelfont=bf]{caption}
\usepackage{subcaption} 
\usepackage{enumitem}
\usepackage{amsmath}
\usepackage{amssymb}
\usepackage{bm}
\setlist{nosep}
\usepackage{authblk}
\usepackage{comment}
\def\M{{Merge}}

\usepackage{color}
\definecolor{deepblue}{rgb}{0,0,0.5}
\definecolor{deepred}{rgb}{0.6,0,0}
\definecolor{deepgreen}{rgb}{0,0.5,0}

\usepackage{listings}

\newcommand\pythonstyle{\lstset{
		language=Python,
		\breaklines=true,
		basicstyle=\ttm,
		otherkeywords={self},             
		keywordstyle=\ttb\color{deepblue},
		emph={MyClass,name,__init__},          
		emphstyle=\ttb\color{deepred},    
		stringstyle=\color{deepgreen},
		frame=tb,                         
		showstringspaces=false            %
}}

\lstnewenvironment{python}[1][]
{
	\pythonstyle
	\lstset{#1}
}
{}

\newcommand\pythoninline[1]{{\pythonstyle\lstinline!#1!}}
\begin{document}
\date{} 

\title{\Large \bf \M: An Architecture for Interconnected Testbed Ecosystems}


\author[1]{Ryan Goodfellow}
\author[1]{Lincoln Thurlow}
\author[1,2]{Srivatsan Ravi}

\affil[1]{University of Southern California, Information Sciences Institute}
\affil[2]{University of Southern California, Dept. of Computer Science}

\maketitle
\subsection*{Abstract}
In the cybersecurity research community, there is no one-size-fits-all solution for \emph{merging} large numbers of heterogeneous resources and experimentation capabilities from disparate specialized testbeds into integrated experiments. The current landscape for cyber-experimentation is diverse, encompassing many fields including critical infrastructure, enterprise IT, cyber-physical systems, cellular networks, automotive platforms, IoT and industrial control systems. Existing federated testbeds are constricted in design to predefined domains of applicability, lacking the systematic ability to integrate the burgeoning number of heterogeneous devices or tools that enable their effective use for experimentation. We have developed the Merge architecture to dynamically integrate disparate testbeds in a logically centralized way that allows researchers to effectively discover, and use the resources and capabilities provided the by evolving ecosystem of distributed testbeds for the development of rigorous and high-fidelity cybersecurity experiments.


%
\section{Introduction}

Cyberspace is now a distinct plane of existence for almost all aspects of life. The complexity, heterogeneity and interconnectivity of that space is increasing rapidly. Cloud computing has given rise to computer clusters the size of sports stadiums that effectively keep tabs on the state of the world, from the systems that hold confidential enterprise information to our personal Dropbox data. The Internet of Things (IoT) movement seeks to saturate our homes and public spaces with small yet highly complex devices that are capable of observing the most private aspects of our lives. Our mobile devices and the back-end systems they communicate with know more about us than we do. The most critical infrastructure on which our quality of life depends is computer controlled, from hospitals to the power grid.

Each of these spaces is rapidly evolving, and the complexity of the computer systems within each space is expanding. Moreover, the growing interconnectivity between these spaces means that experimentally evaluating security and privacy at a \emph{systems level} requires the integration of experimental substrates capable of accurately representing each technological domain. The principle motivation of Merge is to be able to conduct this sort of systems-level cybersecurity experimentation. The major driving factor behind the architectural design intended to satisfy that motivation, is the recognition that the massive complexity associated with each technological domain is best handled by sophisticated domain-specific testbeds.

Merge is designed to interconnect testbeds into an \emph{evolving} ecosystem, capable of supporting rigorous system-level cybersecurity experimentation for the \emph{evolving} needs of the cybersecurity research community.
We summarize the contributions of Merge in the following list that binds architectural elements to experimentation benefit.
\begin{enumerate}[label=(\roman*)]
\item \textbf{Constraint-based expression model}: \emph{define experiments with semantics that directly capture validity.}
\item \textbf{Discovery and realization services}: \emph{find and allocate resources that fall within validity region of a constraint-based experiment.}
\item \textbf{Materialization services}: \emph{orchestrates the provisioning of an experiment across multiple testbed sites.}
\item \textbf{Driver abstraction model}: \emph{incorporate new types of devices dynamically.}
\item \textbf{Resource commissioning model}: \emph{give resource providers proactive control over resources and give researchers instant global visibility of the resource space.}
\end{enumerate}

The primary previous work in this space - developing infrastructure to interconnect testbeds for integrated experimentation, is GENI \cite{geni}. Contributions (i) and (ii) make significant advancements in expressiveness and resource allocation strategies; the design and implementation of (iii) is based on modern fault-tolerance practices that are more resilient and self healing at the infrastructure level; (iv) allows for control plane that dynamically incorporates new devices exposing the full functionality those devices in an end-to-end manner, but captures the complexity of doing so at the device; (v) provides a flexible modeling approach for resource providers to specify resource composition and capability with varying levels of fidelity and instant global visibility without asking each site about available resources through the use of global system snapshots.

One of the primary driving forces behind the design and implementation of Merge is the DHS Cybersecurity Experimentation of the Future (CEF) program \cite{cef}. The ideas from the broader testbed community brought together through the workshops of this program have informed many of the design decisions of Merge.

\section{Outline}
The structure of this paper is organized as follows. 
In Section~\ref{arch} we provide an overview of the Merge architecture, presenting the design from the perspective of the experimenter and the resource provider. 
In Section~\ref{core} we discuss the core components of Merge: the resource representation called \emph{XIR}, the core services \emph{discovery}, \emph{realize}, \emph{materialize} and \emph{commission} and the role of constraints in experiment representation and resource specification.
We refer to each resource provider in a Merge ecosystem as a \emph{site}, in Section \ref{site} we discuss the elements that make up a site.
In Section \ref{prev} we discuss related and prior works which contributed to the overall design of the Merge architecture. We follow up in Section \ref{futu} with our future work for integrating Merge with resource providers, and finally we give concluding remarks in Section \ref{conc}.



%
\begin{figure*}[ht]
\begin{subfigure}{.5\textwidth}
  \centering
  \includegraphics[width=.8\textwidth, height=.25\textheight]{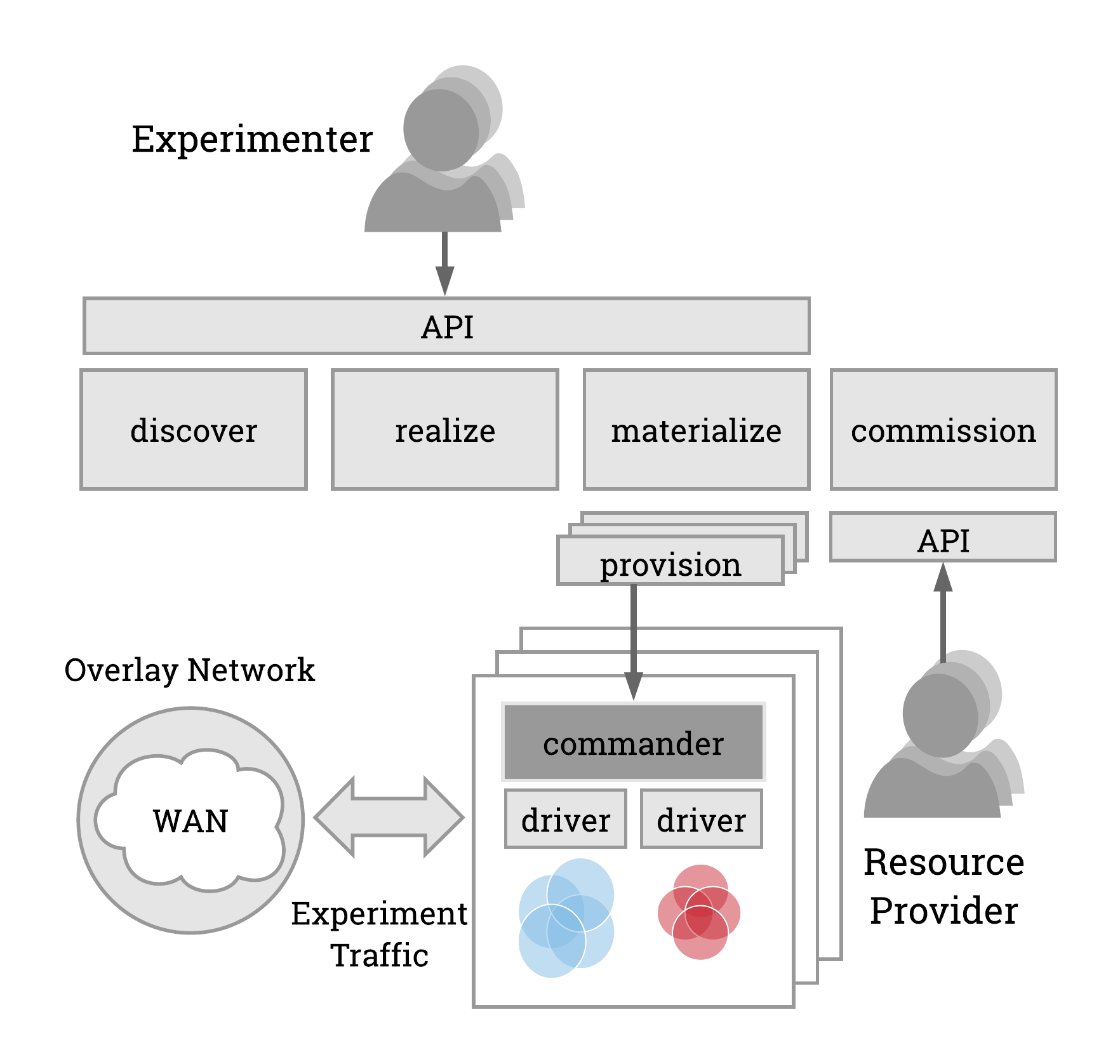}
    \caption{The Merge Architecture}
    \label{merge-arch}
\end{subfigure} %
\begin{subfigure}{.5\textwidth}
  \centering
  \includegraphics[width=.8\textwidth, height=.25\textheight]{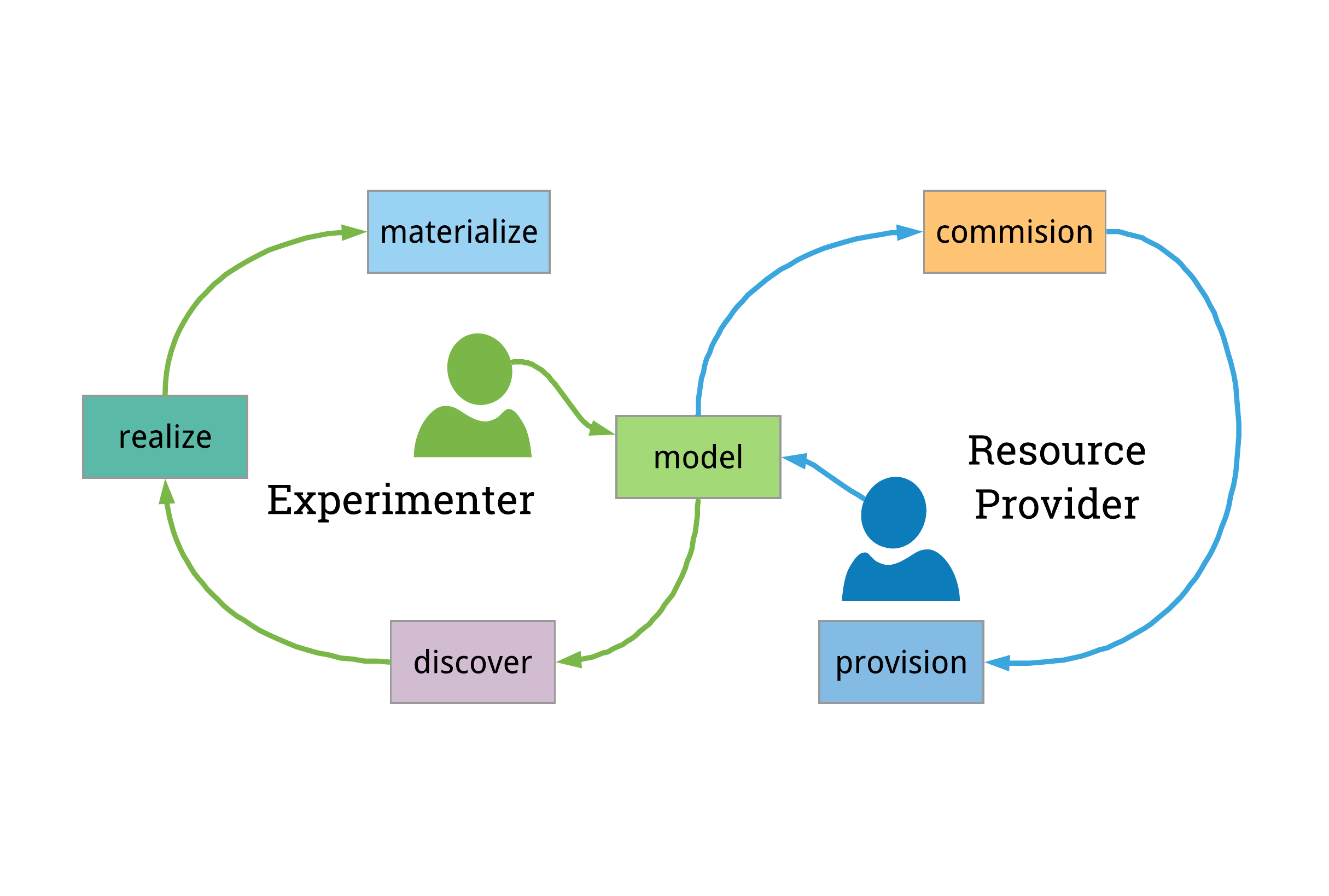}
  \caption{The Merge Experiment Lifecycle}
  \label{merge-life}
\end{subfigure} %
\caption{Merge Architecture and Workflow}
\end{figure*}

\section{Overview}
\label{arch}

The Merge architecture is designed to connect the cybersecurity research, testbed engineering, and compute platform development communities.
The motivation for doing so is to enable a robust and rigorous scientific process for advancing the security of future networked systems relative to the evolution of technologies.
In order to achieve this, the Merge architecture is built around a set of APIs with these communities in mind.
The APIs are partitioned into two categories, one for experimenters and another for resource providers, this partitioning is depicted in Figure~\ref{merge-arch}.
We consider both the testbed engineering and resource development communities to fall under the \emph{resource provider} umbrella.
The central goal of the experimenter API is to allow researchers to \emph{discover} resources that are relevant to their experimentation needs, \emph{realize} experiment models using those resources, \emph{materialize} instantiations of experiments by provisioning acquired resources and finally \emph{execute} behavioral models within a materialized experiment.

The objects of primary concern to experimenters are models that describe the structure and behavior of an experiment - and the artifacts produced by materializing and executing such models.
For resource providers, the objects of primary concern are models that describe the composition and state evolution over a set of resources.
For experimentation to take place at the nexus of these objects, an operating system (in the general sense of the term) is needed to manage experiments, resources and their interactions.
Moreover, for the system to remain relevant in a constantly evolving landscape, its architecture must make very few assumptions about the nature of the technologies in play, the interaction models that exist between them, or the experiments that researchers will build around them.

A good architecture has the capability to evolve with its users. To that end, one of the primary elements of Merge is a scalable representation for experiments and resource models\footnote{We use the word scalable in relation to a representation in the same way the Scala programming language considers its language to be scalable, the language (representation in our case) is able to evolve with its users.}.
We call this representation the eXperimentation Intermediate Representation (XIR).
XIR is the language spoken by the external APIs that are used by both researchers and resource providers. It is also the language spoken by the distributed set of modular services that comprise the core of the Merge architecture.
XIR is a schema-less representation that provides the foundation of Merge's indifference to the form and function of a resource.
The Merge architecture is organized around four core services:
\emph{discovery} - finding available resources,
\emph{realization} - mapping experiment models on to available resources,
\emph{materialization} - turning realizations into reality by provisioning resources, and
\emph{commissioning} - adding and removing resources from the system.
The organization of Merge around these services is depicted in Figure~\ref{merge-arch}, and Figure~\ref{merge-life} depicts how the core services are used in the experimentation life cycle. Experimenters design experiments through an iterative process of developing a model, discovering the resources available to them, selecting a subset of those resources for experiment realization, and finally materializing those resources into an integrated and fully functional experiment environment. 

In Merge experiments are defined as a set of constraints, so there are many possible realizations (set of resources) for any given experiment.
The set of constraints an experimenter starts with in the experiment design process, may not necessarily produce the realization they had intended.
To make the interplay between experiment design and realization effective, realization must be quick.
A distinguishing feature of Merge is the separation of realization and materialization into two distinct activities.
For example, realization is asking for a computation to be made that embeds ones experiment onto a set of available resources and materialization is undertaking the provisioning of the embedding.
The former is relatively quick, even though the algorithmic problem is known to be NP-Hard \cite{vni,emulab}, in practice there are good heuristics and generalized algorithms that compute very large and complex embeddings in milliseconds.
The latter process - materialization, involves heavy-weight operations such as imaging computers with new operating systems or the installation of complex software, and is naturally slow.

The separation of these two activities allows an experimenter to quickly and iteratively realize an experiment. If the provided constraints are too loose (realization results do not meet expectations) then it is a signal to the experimenter that the set of constraints that define their experiment need to be more explicit. Alternatively, a realization may not even be possible for a given set of constraints. This is a signal to the experimenter that constraints need to be relaxed in order to acquire sufficient resources (as long as the relaxation does not invalidate the experiment).

From the resource providers perspective, interacting with Merge is a matter of modeling, commissioning, decommissioning, and provisioning resources. 
In Merge all resources are explicit and must be modeled by resource providers. 
The quality of discovery and realization experienced by experimenters is directly tied to how well providers describe their underlying resource base. 

Commissioning a set of resources means that those resources are under the control of Merge for the period they are in the commissioned state.
An administrator can still reclaim their resources by decommissioning them at any time they are not actively being used \footnote{Force removal for resources under use is possible, but highly discouraged.}.

Provisioning on the other hand, is a reactionary process from the perspective of the resource provider.
When a commissioned resource is selected as a part of a realization and subsequently materialized by an experimenter, provisioning commands are sent to that resource's site by the materialization service.
These commands are received by a component called the \emph{commander}. There is one commander per site, its purpose is to receive provisioning instructions and delegate them to the appropriate \emph{driver}. A site commander also manages the experiment's traffic flow in and out of a site by dynamically configuring a Software Defined Networking (SDN) component called \emph{Hummingbird} that is described in Section~\ref{hummingbird}.

Drivers are the components that are ultimately responsible for implementing the provisioning required by the materialization process. Drivers are also the abstraction by which the Merge architecture encapsulates the \emph{particular complexity} of the vast array of devices that comprise its resource base.
Just like the classical operating system driver model, there is one driver for each \emph{type} of device.
Each driver conforms to a simple state-machine model that the allows the Merge materialization service to track and transition the state of devices during provisioning.

%
\section{Core Components}
\label{core}

In this section we describe the services that make up the Merge core. 
The core services partition the Merge's functionality into four independent subsystems that communicate with each other through well defined \texttt{protobuf} \cite{protobuf} interfaces.
Each of the core services is stateless.
This makes replication and subsequently fault-tolerance schemes that are based on a replication model easy to implement from a system design point of view.
The services gather their required state from a set of databases. 
The only thing that is required of the databases, is that they support transactions. 
This is necessary because both the realization and materialization services employ optimistic concurrency techniques and require transactions to maintain correctness.
As we describe each of the core services, the modeling representation XIR, as it specifically pertains to each service will be presented in context.

\subsection{Discovery \& Realization}
\label{discover}
\label{realize}




The \textbf{discovery} process computes a map of an experiment network $\bm{x}$ onto a candidate map $\bm{c}$.
\begin{equation}
  \bm{x} := (\bm{n}_x, \bm{l}_x) \mapsto \textbf{c}(t) := (\bm{n}_x, \bm{N}_r(t))
\end{equation}
Here the $\bm{n}_x$ are experiment nodes, $\bm{l}_x$ are experiment links and $\bm{N}_r$ is a superset of resource nodes. When taken together $(\bm{n}_x, \bm{N}_r)$ is a map from each experiment node $n \in \bm{n}_x$ to a set of satisfying resource nodes $N \in \bm{N}_r$.

\begin{figure}[ht]
    \centering
    \includegraphics[width=.95\columnwidth]{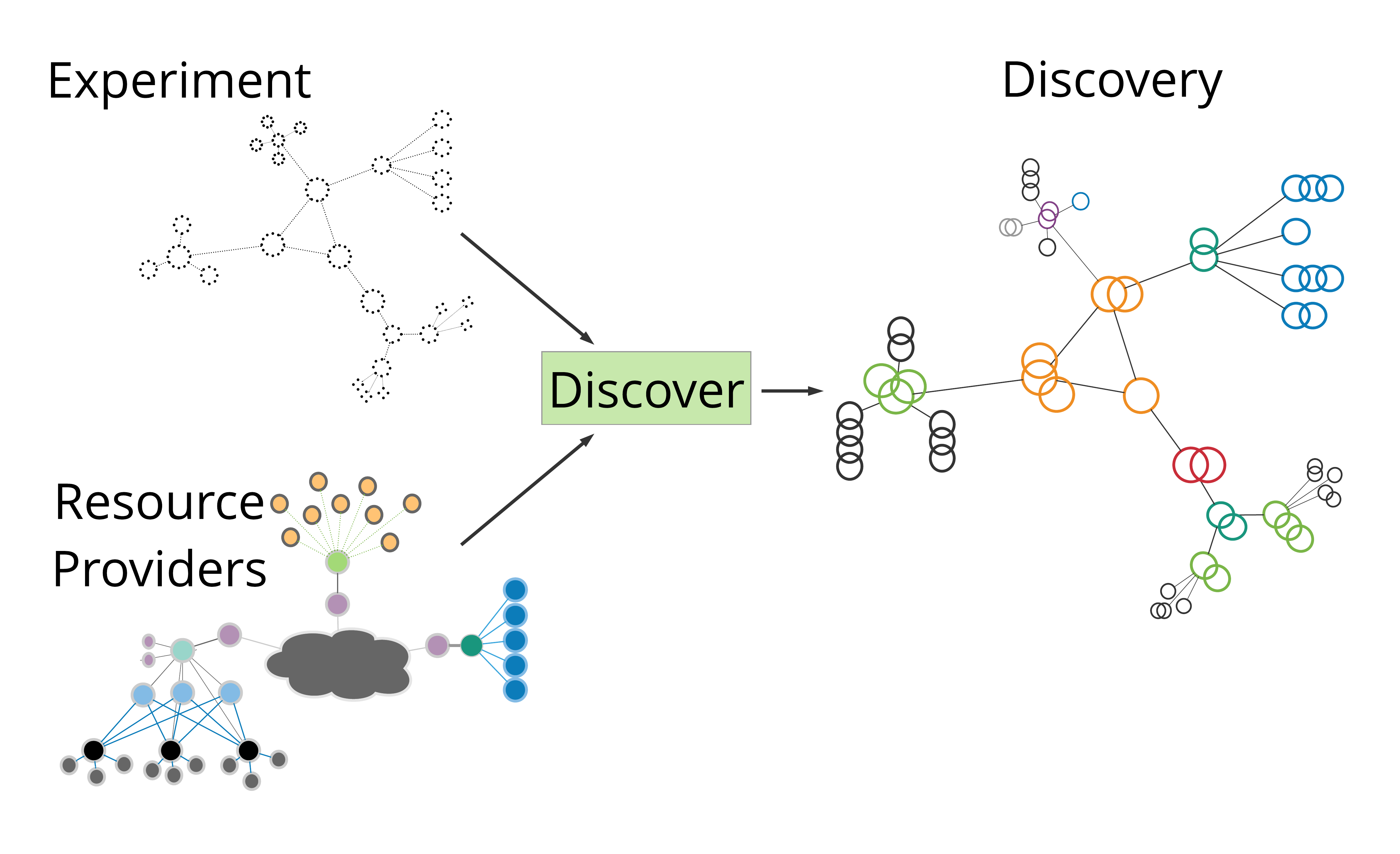}
    \caption{Discovery of resources for an experiment}
    \label{disco}
\end{figure}

This mapping lets the experimenter see what is out there in terms of useful resources for their experiment. The process is depicted graphically in Figure ~\ref{disco}. 


The \textbf{realization} process is very similar to the discovery process. However, instead of mapping an experiment network onto a candidate map, it is mapped onto a resource network $\bm{r}$.
\begin{equation}
  \bm{x} := (\bm{n}_x, \bm{l}_x) \mapsto \textbf{r} := (\bm{n}_r, \bm{l}_r)
\end{equation}
Here the $\bm{n}_r$ are resource nodes and $\bm{l}_r$ are resource links. The graphical depiction is the same as Figure~\ref{disco} except that instead of getting a puffed up graph where the nodes are candidate sets, a standard graph is returned where each node is the chosen resource to underpin an experiment element.

One of the principle features Merge aims to provide is interactive realization.
What this means from a design standpoint is that the algorithms that implement realization need to be fast and reliable.
One could imagine an architecture for a system that accomplishes the same things as Merge, but is based on a fully distributed design.
In this case the implementation of the core services would require fully distributed locks.
In particular, every realization request would require a mutual exclusion envelope that includes at least two global snapshots, one for the state sufficient to run the embedding algorithm and subsequently another one for the reservation of the selected resources pending a successful embedding computation.
The envelope could be implemented pessimistically with a lock, which goes scale-fail very quickly (for obvious lock the world reasons) with a large resource base.

Alternatively if implemented optimistically, transactional rollback and retry under load (another two snapshots per) adds up very fast.
With cross country round trip times anywhere between 50-200 ms and the large number of round trips required to reach a consistent snapshot, the fully distributed approach will be neither fast nor reliable.
A Paxos or Raft-like \cite{paxos,raft} protocol supporting dynamic membership would have to be implemented across the entire network.
Alternatively, limiting the dispersion of critical state to a scale out environment with high-performance databases and reliable fast networks specifically designed to take on this sort of concurrent workload at scale seems a much better choice.

Owing to the scalable representation model on which Merge is based, the realization engine is also scalable in the sense that it naturally evolves along with experiments and the resource base.
The realization and discovery engines operate over XIR representations.
The one assumption XIR makes about the models it represents is that the structure is that of a network. 
XIR has the semantics of a network (in the connected elements sense not the data network sense) cooked into the representation.
Every node and link in an XIR network has a set of properties associated with it.
These properties manifest themselves in one of two ways depending on what is being represented.
If an experiment element is being represented, these properties are fuzzy in that they are represented by constraints.
If an resource element is being represented the properties are concrete.

As a simple example consider a model of an IoT network that may be part of an experiment as shown in Listing \ref{iot-code}, showcasing how properties are able to represented as constraints in XIR.
This model uses the Python language bindings for XIR.
Looking at the interconnect between the \texttt{breaker} and \texttt{xbeehub} nodes we see that the experimenter requires that the communications stack be \texttt{zigbee}, the bandwidth is no greater than \texttt{100 kbps} and the loss rate on the link is greater than \texttt{5\%}.
Furthermore, XIR allows experimenters to specify requirements through key-values (such as operating system, or memory).


\begin{figure}[!t]
	\vspace{-\topsep}
	\begin{lstlisting}[captionpos=b,label=iot-code,caption=Describing a topology in Python XIR,basicstyle=\scriptsize\ttfamily]
breaker = topo.node({
    'name': 'breaker',
    'image': select('riot'),
    'memory': { 'capacity': gt(xir.mB(256)) },
})
xbeehub = topo.node({
    'name': 'xbeehub',
    'image': choice([
        select('debian-9'),
        select('ubuntu-snap')
    ])
})
topo.connect([breaker, xbeehub], {
  'stack': eq('zigbee'),
  'bandwidth': lt(xir.kbps(100)),
  'loss': gt(xir.percent(5))
})
	
	\end{lstlisting}
	\vspace{-1em}
\captionsetup[listing]{position=below,skip=-10pt, justification=centering}
\end{figure}

The job of the realization and discovery engines is one of matching.
They understand a standard set of constraint operators and how to match the data types bound by those operators to concrete values in resource specifications.
Like every other core component, realization and discovery are tied to well defined interfaces.
This means that the underlying implementation can be changed easily or there may be multiple implementations active at a time and users can specify which they would prefer.
Currently, we have two implementations of the realization engine.
One is a simple greedy heuristic like the one presented as the base model in \cite{vni}.
The other is based on an off-the-shelf general satisfiability (SAT) solver, PicoSAT \cite{picosat}.



\subsection{Materialization}
\label{materialize}

Materialization is the process by which Merge transitions an experiment from the realized to materialized state.
When an experiment has been realized, it means that the resources necessary to instantiate it have been computed and allocated.
Materialization is the process of provisioning the resources selected in the realization phase according to the data specified in an experiment model.
Provisioning a resource according to data in an experiment model can mean a wide variety of things.
Consider two very different examples: an experiment specifies a node has a particular operating system, and an experiment specifies that an industrial control device is configured with a referenced IEC-61131 program.
The materialization engine accommodates both of these use cases, but knows nothing of operating systems or IEC-61131.
In fact the materialization has very little to do with provisioning, all of the real work of provisioning is delegated to drivers, which are described later in Section~\ref{site}.
The primary goal of the materialization engine is to orchestrate provisioning across a distributed network of resources.

To carry out provisioning over a distributed network, we have modeled the materialization engine after the Kubernetes \cite{kube} state convergence model, which solves a similar problem.
When a materialization for an experiment is started, every component of that experiment gets an entry in a data structure called a \emph{stateboard}.
The stateboard is a 2-level (\texttt{experiment-id}, \texttt{resource-id}) lookup table that maps an unique identifiers for an experiment resource to the associated materialization state.


Each entry in the stateboard starts at the \texttt{zero} state, with a target state of \texttt{configured}.
The state transitions in between will be discussed in Section~\ref{driver}. It is the materialization engines responsibility to drive each elements state from \texttt{current} to \texttt{target}.
It accomplishes this through a set of agents that are continually monitoring the stateboard.
When an agent finds an entry where the current state is not equal to the target state, it tries to acquire a lease that gives it the right to claim responsibility for driving the \texttt{current} state to the \texttt{target} state.
Architecturally, claiming a lease is required to be an atomic transaction, and the state of each entry must be guarded by a claim.
This allows the agents to scale, and is sufficiently general to allow for swapping out the underlying data store technology if desired.

When it comes to imparting a provisioning action on a resource, the materialization agents work via delegation.
They collect the information from the experiment database that is required to provision a particular resource and then pass that information on to the driver responsible for performing the provisioning.
The materialization agent does not know which driver to contact.
However, it does know what resource provider's site a resource is associated with, and how to look up the model for that site in the site's data store.
Each site model includes a fully qualified domain name that the site commander can be contacted at.
The commander acts as a proxy between the materialization agents running in the core and the various resource drivers running within sites.
Thus the job of the materialization agent is one of ensuring state convergence by delegation.


\subsection{Commissioning}

\begin{figure}[ht]
\begin{subfigure}{\columnwidth}
  \centering
  \includegraphics[width=.95\columnwidth]{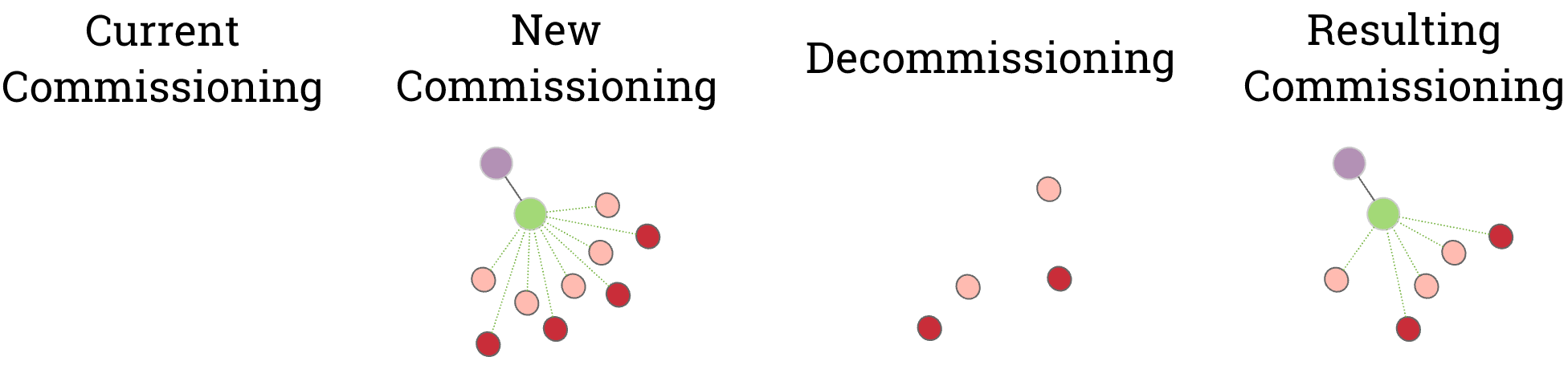}
    \caption{Commissioning - Simple Case}
    \label{commission-simple}
\end{subfigure} %
\begin{subfigure}{\columnwidth}
  \centering
  \includegraphics[width=.95\columnwidth]{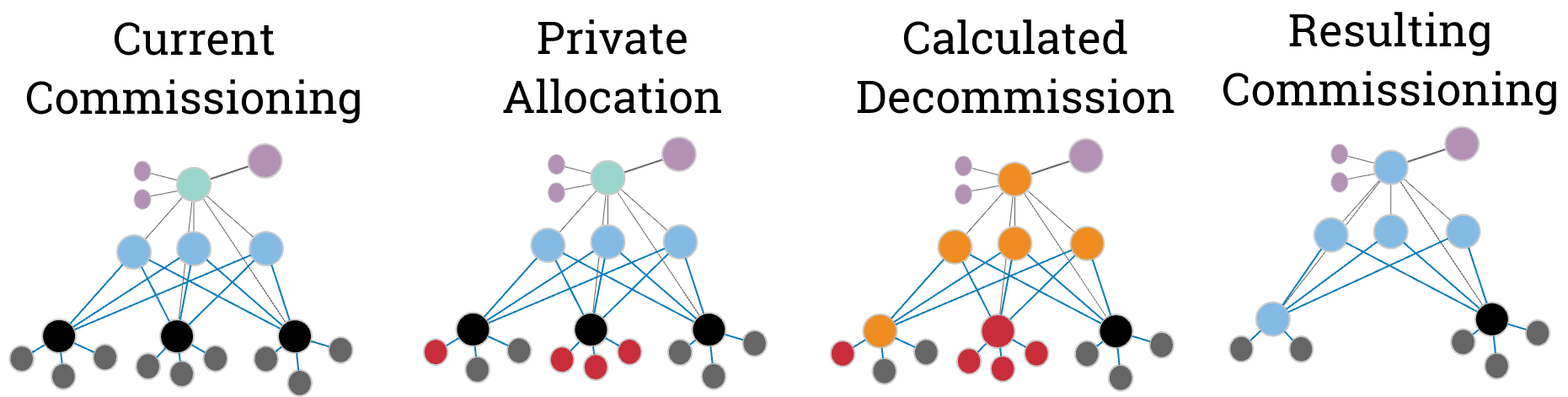}
  \caption{Commissioning - Fragmented Case}
  \label{commission-fragmented}
\end{subfigure} %
\caption{Commissioning simple and complex XIR networks}
\label{commission}
\end{figure}

Commissioning is the process by which resources are added to and removed from a Merge managed ecosystem.
The unit of commissioning and decommissioning is an XIR network.
This includes everything from a single node network to a complex interconnected fragment of a resource provider's switching mesh.
Figure~\ref{commission} shows the two basic modes of commissioning and decommissioning, \emph{simple} and \emph{fragmented}.

In the simple case, resources can be commissioned and decommissioned atomically. Consider the example of Figure~\ref{commission-simple}.
In this case the initial commissioning starts out empty.
Then a network is added in a single commissioning request, later the operator of that site decides to take a few nodes away from Merge's control.
This decommission request involves four nodes.
The resulting network is shown as the final diagram on the far right of Figure~\ref{commission-simple}.
This case is considered simple because there are no resources that are impacted besides the ones being commissioned or decommissioned.

In Figure~\ref{commission-fragmented}, we consider the slightly more complex case of fragmented commissioning.
In this example we start with an already commissioned set of resources.
Then the red-highlighted nodes are removed from the commissioning for external use by the site administrator.
In this case, the nodes are a part of a folded Clos network and the removal of nodes has an impact on other elements in the network model i.e., the external use of these nodes will continue to consume resources throughout the network topology. 
In particular the backplanes of the switching elements highlighted in orange will be reduced, but in a way that is invisible to the Merge resource allocation mechanisms. 
Thus the model must be modified to account for this loss in bandwidth.

In the core, Merge simply understands resource addition and subtraction in a very abstract sense.
It has no intelligence about how network capacity properties propagate across a network topology.
It is the responsibility of site administrators to calculate accurate networks as input to the commissioning and decommissioning APIs.

 \section{Site Components}
\label{site}

There are three kinds of components that make up resource provider site: a commander, drivers for resources, and the Hummingbird virtual network.
The commander of a site serves the purpose of: managing drivers, providing a local commissioning interface to site administrators, and managing the Hummingbird virtual network.
The commander is the interface for a resource provider to interact with Merge and vise versa.
Drivers are the main component of a resource provider, they are responsible for implementing the materialization agent's actions on a resource.  
Thus, the control plane for a resource as perceived by Merge, is implemented in the driver for that resource.
Finally, Hummingbird is responsible for creating isolated experiment networks between resource providers across wide-area networks. It is not assumed in advanced what the functionality of a network resource is for initiating isolated connections, Hummingbird provides isolation translation services across a variable set of resources.

\subsection{Commander}

The commander is the interface between the resources and Merge.
Resources are added to Merge through the commander.
Resource providers commission resources through the commander, which generates a psuedorandom 128 bit UUID for each resource that Merge will validate before accepting.
The commander also keeps enough state to map device UUIDs to driver instances.
When a command from a materialization agent is received, the commander looks up the UUID for the resource, and forwards the commands to the associated driver, specifying the UUID in the case where driver is responsible for multiple resources.
When an experiment has completed, Merge engages in a dematerialization protocol with the site commander with the result being that all resources associated with an experiment at that site, both virtual and physical are freed.

\subsection{Driver}\label{driver}

In the Merge architecture, drivers are responsible for resource's actions according to experiment specifications.
They need not worry about where to get the resource specifications, that is the job of the materialization agents in the core.

\begin{figure}[ht]
    \centering
    \includegraphics[width=.65\columnwidth]{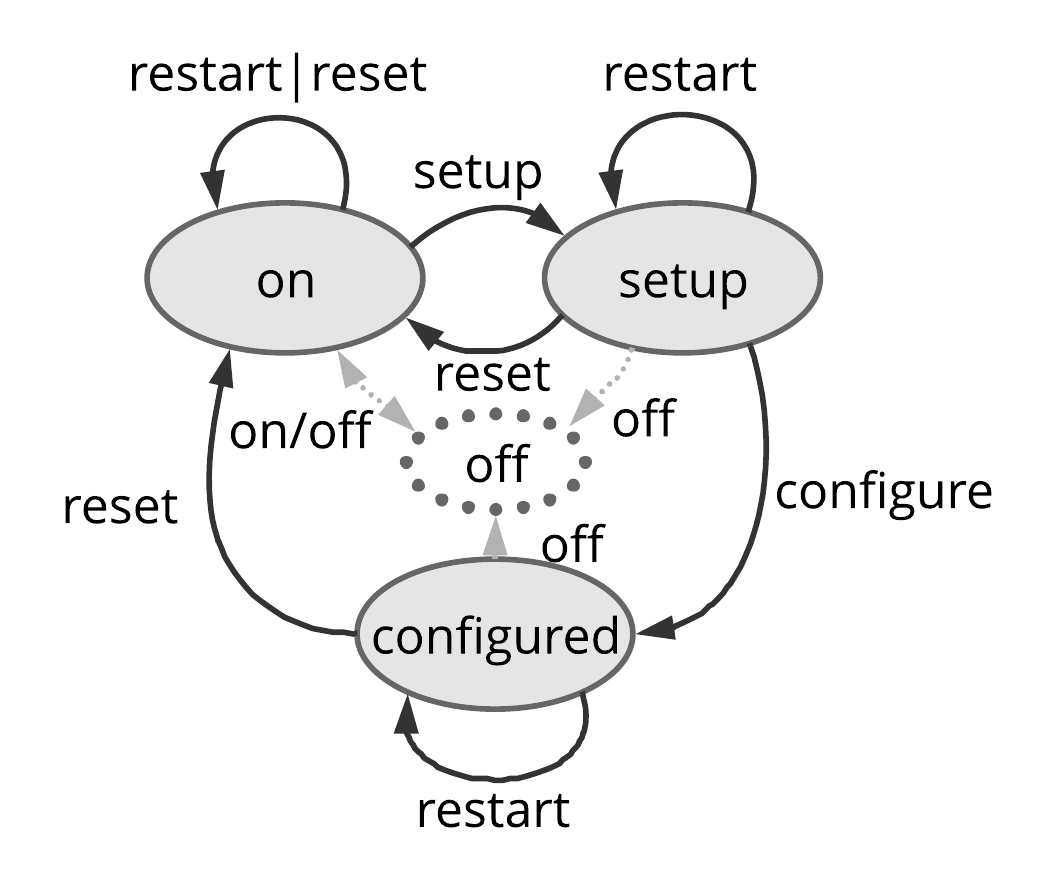}
    \caption{Merge's driver state machine for resources}
    \label{driversm}
\end{figure}

Resources in Merge are explicit \emph{things}, they can be networking ports, circuit breakers, FPGAs, etc.
A driver may control one or more resources.
Drivers are responsible for implementing a common state machine shown in Figure~\ref{driversm}.
The four states: on, off, configure, and setup are all the stages necessary to implement an experiment.
The setup state ensures the resource has been correctly setup. For instance, if the resource is an embedded controller, setup may include the desired firmware to be loaded.
Once in configured state, the resource is ready to be used by Merge.
Merge ensures that all resources have reached a configured state before an experiment reports as materialized. 
Finally, the on and off states are initial states meant to be used during commissioning and decommissioning to enforce fresh state.


Anyone can write a driver for a Merge resource, but like drivers that are written for devices that get plugged into a computer, it makes the most sense for the people working directly on those devices to implement the functionality.
In the operating systems space, developers of devices write drivers because they want their customers to be able to use their devices with the operating system in question.
The same benefit model can be applied to Merge. If device developers can be incentivized to write Merge drivers for their resources based on the cybersecurity research community having access to their resources for experimentation, everyone benefits.



\subsection{Hummingbird}\label{hummingbird}

Materializing experiments across resources that are distributed over multiple resource provider's networks via the wide-area networks introduces the challenge of maintaining network isolation.
Unique to heterogeneous testbeds is the issue of not being able to make assumptions about the type of network connectivity between devices or the networking components that provide that connectivity.
For instance, a resource's first hop connection to Merge may be through LTE, ZigBee, Ethernet, or fiber-channel connections.
Thus, it is imperative that resource provider sites are able to collectively build isolated overlay networks that do not conflict, interfere, or impose artificial constraints on the guest networks.  

To accomplish globally isolated networks composed of local networks with disparate isolation mechanisms, we are developing a technology called Hummingbird.
Hummingbird is a three-tired architecture, the tiers consist of resources (IoT, servers, etc) at the lowest level, connected to devices which can support isolation (VLAN, CDMA channels, Zigbee CID), followed by Hummingbird nodes which connect to wide-area networks (WAN).
For experiment traffic leaving a site, the Hummingbird node translates the downstream (toward the site) isolation mechanism into VXLAN for transport across the WAN.
For experiment traffic entering a site, the Hummingbird node translates upstream (toward the Internet) VXLAN packets into the appropriate isolation mechanism for the target network within the site.



\section{Previous Work}
\label{prev}

Testbed federation is a well-explored topic, the focus of which has been on how to best share resources in a manner that scales and is beneficial both to resource providers and experimenters.

The National Science Foundation's GENI project \cite{geni} was the first such large effort push for creating a framework to allow testbeds across the United States to interconnect and share their resources.  Resource providers, mainly testbeds, join the federation, where each resource provider maintains control over the resources they provide. GENI uses Rspec \cite{rspec} for describing experiment resources, an XML-based language. Merge is closer to PlanetLab \cite{planetlab} in design by having centralized control of resources. The experimenter side of Merge's XIR is constraint based and schema-less as opposed to an XML schema that deals only with concretely specifying particular resources. 

PlanetLab as precursor to GENI provided a global research platform. Experimenters would reserve slices based on the SFA \cite{sfa} of the topology for each experiment. In contrast, Merge does not use slices, and the resource specifications extend beyond x86 machines running Linux to a breadth of devices running whatever operating system is supported by the device and site administrator.

The FIRE initiative \cite{fire} has brought about multiple IoT testbeds with diverse resources and application usages. The SmartSantander \cite{smartsantander} testbed brings IoT to the city.  SmartSantader currently supports representational state transfer (REST) requests to allocate IoT devices for experimenters. The FIT loT-LAB \cite{iotlab} is smaller in size compared to SmartSantander, however it provides additional device support and programmability, including specifying operating systems. However, compared to Merge's microservice oriented core, IoT-Lab has a monolithic core which makes extensibility and maintainability more difficult. 

The SAVI \cite{savi} architecture supports similar device control to Merge.  In SAVI, every resource has a controller and the controller implements the actions of the device.  However the SAVI infrastructure assumes a directly connected infrastructure and a single administrative domain of control.


%
\section{Future Work}
\label{futu}
We are at the beginning stages of helping a few testbed providers in joining Merge.
One of the conditions of providing resources from testbed operators was the ability for the testbed to implement constraints on the type of experiment that could be run over their resources.
This is the problem of governance, we are working on developing a simple scheme to allow resource providers the ability to dictate to Merge the level of risk a resource provider is willing to take on for a cybersecurity experiment.

The Hummingbird SDN edge for sites is under active development. A few common isolation protocols have been implemented such as VLAN and VXLAN, but to be pragmatic there are many more under development.

%
\section{Conclusion}
\label{conc}
We have presented the Merge architecture, whose motivation is to enable systems level experimentation for cybersecurity over a wide range of technological domains. 
Through the use of XIR, a constraint-based realization engine and the driver-model, Merge is able to be agnostic to the underlying resources, while also being able to harness the particular sophistication of the resource.
By modularizing the main components of the architecture, and explicitly separating realization and materialization, Merge allows experimenters to have more flexibility and control through interactive and iterative experimentation.

%
\section*{Acknowledgements}
This work was sponsored by the Department of Homeland Security under Contract No. HSHQDC-16-C-00024
and the National Science Foundation under Contract
No. ACI-1346277. Any opinions, findings, and conclusions or recommendations expressed in this material are
those of the author(s) and do not necessarily reflect the
views of the Department of Homeland Security. CEF is an NSF/DHS initiative for building Cyber Experimentation of the Future testbeds. The authors would like to acknowledge Erik Kline, Genevieve
Bartlett and Jim Blythe for thoughtful insight and commentary during the development of this paper.
\section*{Availability}
Merge is currently available on Gitlab. \url{https://gitlab/mergetb}
{\footnotesize \bibliographystyle{acm}
\bibliography{ceftb}}

\end{document}